\documentclass[10 pt, twocolumn, twoside]{IEEEtran}

\usepackage{graphicx,amssymb,amstext,amsmath,color}
\DeclareGraphicsExtensions{.eps,.pdf,.png,.jpg,.gif,.jpeg,.pstex}

\input{epsf.sty}

\usepackage{times, epsf}
\usepackage{amsmath,amssymb}
\usepackage{url}
\usepackage[font = footnotesize]{caption}
\usepackage{ float}
\usepackage{algorithm}
\usepackage{algorithmic}
\usepackage{graphicx}
\usepackage{subfig}
\usepackage{cite}
\usepackage{multirow,tabularx}
\usepackage[pdftex]{epsfig}
\input{epsf.sty}

\usepackage{stackrel}
\usepackage{amsfonts}
\input{epsf.sty}

\usepackage{times, epsf}
\usepackage{graphicx}
\usepackage{cite}
\usepackage{multirow,tabularx}
\usepackage{epstopdf}

\newcommand{\expect}[1]{{\mathbb{E}}\left[{#1}\right]}

\newcommand{\E}[1]{\expect{#1}}

\newcommand{\ba}{\begin{array}}
\newcommand{\ea}{\end{array}}

\def\PARstart#1#2{\begingroup\def\par{\endgraf\endgroup\lineskiplimit=0pt}
    \setbox2=\hbox{\uppercase{#2} }\newdimen\tmpht \tmpht \ht2
    \advance\tmpht by \baselineskip\font\hhuge=cmr10 at \tmpht
    \setbox1=\hbox{{\hhuge #1}}
    \count7=\tmpht \count8=\ht1\divide\count8 by 1000 \divide\count7 by\count8
    \tmpht=.001\tmpht\multiply\tmpht by \count7\font\hhuge=cmr10 at \tmpht
    \setbox1=\hbox{{\hhuge #1}} \noindent \hangindent1.05\wd1
    \hangafter=-2 {\hskip-\hangindent \lower1\ht1\hbox{\raise1.0\ht2\copy1}%
    \kern-0\wd1}\copy2\lineskiplimit=-1000pt}

\newcommand{\bit}{\begin{itemize}}
\newcommand{\eit}{\end{itemize}}

\newcommand{\twopartdef}[4]
{
	\left\{
		\begin{array}{ll}
			#1 & \mbox{if } #2 \\
			#3 & \mbox{if } #4 
		\end{array}
	\right.
}

\long\def\comment#1{}

\newfont{\bbb}{msbm10 scaled 700}

\newfont{\bb}{msbm10 scaled 1100}

\newcommand{\Bm}{{\bf B}}
\newcommand{\Cm}{{\bf C}}
\newcommand{\Dm}{{\bf D}}

\newcommand{\Gm}{{\bf G}}
\newcommand{\Hm}{{\bf H}}

\newcommand{\Sm}{{\bf S}}

\newcommand{\Ec}{{\cal E}}
\newcommand{\Fc}{{\cal F}}
\newcommand{\Gc}{{\cal G}}
\newcommand{\Hc}{{\cal H}}

\newcommand{\Kc}{{\cal K}}

\newcommand{\Nc}{{\cal N}}

\newcommand{\Uc}{{\cal U}}

\newcommand{\gammav}{\hbox{\boldmath$\gamma$}}

\newcommand{\muv}{\hbox{\boldmath$\mu$}}

\newcommand{\Thetav}{\hbox{\boldmath$\Theta$}}

\newcommand{\omegav}{\hbox{\boldmath$\omega$}}

\renewcommand{\arg}{{\hbox{arg}}}

\newcommand{\transp}{{\sf T}}

\begin{document}
\title{Adaptive Video Streaming in MU-MIMO Networks}
\author{\IEEEauthorblockN{D. Bethanabhotla, G. Caire and M. J. Neely}
}
\maketitle

\begin{abstract} 
We consider extensions and improvements on our previous work~\cite{bethanabhotla2013joint} on dynamic adaptive video streaming in a multi-cell multiuser ``small cell'' wireless network. In~\cite{bethanabhotla2013joint} we treated the case of single-antenna base stations and, starting from a network utility maximization (NUM) formulation, we devised a ``push'' scheduling policy, where users place requests to sequential video chunks to possibly different base stations with adaptive video quality, and base stations schedule their downlink transmissions in order to stabilize their transmission queues. In this paper we consider a ``pull'' strategy, where every user maintains a request queue, such that users keep track of the video chunks that are effectively delivered. The pull scheme allows to download the chunks in the playback order without skipping or missing them. In addition, motivated by the recent/forthcoming progress in small cell networks (e.g., in wave-2 of the recent IEEE 802.11ac standard), we extend our dynamic streaming approach to the case of base stations capable of multiuser MIMO downlink, i.e., serving multiple users on the same time-frequency slot by spatial multiplexing. By exploiting the ``channel hardening'' effect of high dimensional MIMO channels, we devise a low complexity user selection scheme to solve the underlying max-weighted rate scheduling, which can be easily implemented and runs independently at each base station.  Through simulations, we show MIMO gains in terms of video streaming QoE metrics like the pre-buffering and re-buffering times.
 \end{abstract}

\section{Introduction}  \label{sec:intro-dilip}

Wireless data traffic is predicted to increase dramatically in the next few years, up to two orders of magnitude by 2020 \cite{cisco66}. This increase is mainly due to on-demand video streaming, enabled by multimedia devices such as tablets and smartphones. It is well understood that the current trend of cellular technology (e.g., LTE~\cite{sesia-LTE}) cannot cope with such traffic increase, unless the density of the deployed wireless infrastructure is increased correspondingly. This motivates the recent flurry of research on massive and dense deployment of base station antennas, either in the form of ``massive MIMO'' solutions (hundreds of antennas at each cell site \cite{hoydis2011massive}) or in the form of very dense small-cell networks \cite{ hoydis2011green}. While discussing the relative merits of these approaches is out of the scope of this paper, we mention here that the `small-cell MU-MIMO' solution seems to be most immediately applicable, since it can leverage the rapidly evolving MIMO technology advances of wireless local-area networks standards, such as IEEE 802.11 ac with MU-MIMO capability~\cite{802.11ac}.
This paper focuses on the problem of dynamic adaptive video streaming in a  wireless network formed by a number of small-cell base station (helper) nodes, employing MU-MIMO technology,
serving multiple wireless users over a given  geographic coverage area and on the same shared channel bandwidth. 

{\bf Contributions:}  We introduce the notion of a {\it request queue}. This is a virtual queue, maintained by each user, that serves to sequentially request video chunks from helper nodes, such that the choice of the helper node and the quality at which each video chunk is requested can be adaptively adjusted. In order to obtain a dynamic policy with provable optimality properties, we formulate a NUM problem~\cite{yi2008stochastic} and solve it using the drift plus penalty (DPP) approach in the framework of  Lyapunov Optimization~\cite{neely2010stochastic}. The obtained solution is provably asymptotically optimal in a per-sample path sense (i.e.,  without assuming stationarity and ergodicity of the underlying network state process \cite{neely2010stochastic}). 
Furthermore, it naturally decomposes into two sub-policies: ``congestion control" which is implemented independently at every user and, ``transmission scheduling" which is implemented independently at every helper. The congestion control decision consists of each user 
adaptively selecting the video quality level of the chunks and ``virtually" placing them in its request queue. Note that this does not mean the user already has the chunk, but the chunk is ``virtually" placed in the request queue and will be taken out when it is effectively delivered to the user. In addition, the user broadcasts its request queue length to the helpers in its current vicinity and requests from them only those bits which are at the `head of line' of its queue.  In this way, the user always downloads chunks in the playback order and does not skip any of them.
The transmission scheduling decision consists of each ``MU-MIMO" base station greedily choosing the subset of users to be served by multiuser MIMO spatial multiplexing based on the request queue length information broadcasted by the users in its vicinity.



\section{System Model} \label{sec:sysmodel-dilip}
We consider a discrete, time-slotted wireless network with multiple users and multiple helper stations sharing the same bandwidth. The network is defined by a bipartite graph $\Gc = (\Uc, \Hc, \Ec)$, where $\Uc$ denotes the set of users, $\Hc$ denotes the set of helpers, and $\Ec$ contains edges for all pairs $(h,u)$ such that there exists a potential transmission link between $h \in \Hc$ and $u \in \Uc$. We denote by $\Nc(u) \subseteq \Hc$ the neighborhood of user $u$, i.e., $\Nc(u) = \{ h \in \Hc : (h,u) \in \Ec\}$. Similarly,  $\Nc(h) = \{u \in \Uc : (h,u) \in \Ec\}$.  
Each user $u \in \Uc$ requests a video file $f_u$ from a library $\Fc$ of possible files. 
Each video file is formed by a sequence of  chunks. Each chunk corresponds to a group of pictures (GOP) that are encoded and decoded 
as stand-alone units~\cite{sanchez2011idash}. Chunks have a fixed playback duration, 
given by $T_{\rm gop} = \mbox{(\# frames per GOP)}/\eta$, where $\eta$ is the frame rate, expressed in frames per second. The streaming process consists of transferring chunks from the helpers to the requesting users such that the playback buffer at each user contains the required chunks at the beginning of each chunk playback deadline. The playback starts after a short pre-buffering time, during which the playback buffer is filled by a determined amount of ordered chunks. The details relative to pre-buffering and chunk playback deadlines are discussed in 
Section \ref{sec:prebuffering}.

Each file $f \in \Fc$ is encoded at a finite number of different quality levels $m \in \{1, \ldots, N_f\}$. This is similar to the implementation of several current video streaming 
technologies, such as Microsoft Smooth Streaming and Apple HTTP Live Streaming~\cite{begen2011watching}. Due to the variable bit rate (VBR) nature of video coding\cite{ortega2000variable}, the quality-rate profile of a given file $f$ may vary from chunk to chunk. We let $D_f(m,t)$ and $B_f(m,t)$ denote the video quality measure (e.g., see \cite{wang2004image}) and the number of bits per pixel for file $f$ at chunk time $t$ and quality level  $m$ respectively. 
Letting  $N_{\mathrm{pix}}$ denote the number of pixels per frame, a chunk contains $k = \eta T_{\rm gop} N_{\mathrm{pix}}$ pixels. Hence, the number of bits in the $t$-th chunk of file $f$, encoded at quality level $m$, is given by $k B_f(m,t)$. 

Each user maintains a {\it request queue} $Q_u$ of bits that it wants to download possibly from different helpers it might associate with during its video streaming session. Note that $Q_u$ is different from the actual playback buffer $\Psi_t$ (see Section~\ref{sec:prebuffering}) and acts like a bookkeeper by maintaining the list of chunks that have been requested but not downloaded yet. In addition, each user $u$ independently makes the decision of choosing the quality mode $m_u(t)$ for chunk time $t$. This choice affects the choice of the quality $D_{f_u}(m_u(t),t)$ and the size  
$kB_{f_u}(m_u(t),t)$ of the chunk $t$ that it places in its request queue $Q_u$. The dynamics of $Q_u$ for each user $u$ is given by:
\begin{align}
\label{q-update}
  Q_{u}(t+1)=\max\{Q_{u}(t)-n\mu_{u}(t)+kB_u(t), 0\} \;\;\;\; \forall~ u \in \Uc,
\end{align}
where $n$ denotes the number of physical layer channel symbols corresponding to a time slot of duration $T_{\rm gop}$, and $n\mu_{u}(t) = \sum_{h \in \Nc(u)}n\mu_{hu}(t)$ is the aggregate number of video-encoded bits per time slot that the user is able to download from its neighboring helpers. Note that $\mu_{hu}(t)$ is the channel coding rate (bits/channel symbol) of the 
transmission from helper $h$ to user $u$ at time $t$. Here, we assume that user $u$ at time $t$ can receive $\mu_{u}(t)= \sum_{h \in \Nc(u)}\mu_{hu}(t)$ bits/channel symbol by simultaneously downloading $n\mu_{hu}(t)$ bits from helpers $h$ in $\Nc(u)$. Although this is not implemented in 802.11 networks, receiving multiple data streams from multiple base station is definitely possible (e.g., in CDMA system with macro diversity). In any case, in this paper, we do not assume that the user is able to perform joint or successive decoding of the multiple streams: when a user is served by multiple helpers on the same time slot, each stream is decoded independently and treats everything else as Gaussian noise. 
Note that the $kB_u(t)$ bits that user $u$ places in its request queue $Q_u$ correspond to the chunk $t$ while the $n\mu_{u}(t)$ bits that it downloads from its neighboring helpers correspond to the chunks at the `head of the line' of $Q_u$. In this way, the user can download chunk $t+1$ only after having downloaded chunk $t$ and consequently receive chunks in order of playback. This is in contrast to the `push' scheme in~\cite{bethanabhotla2013joint} where a user, depending on the queue lengths of the neighboring helpers, may download chunks out of order. 
We consider two possible physical layer systems in this paper. PHY A is the system which was already considered in  \cite{bethanabhotla2013joint}, where each helper $h$ has a single antenna and serves its neighboring users $u \in \Nc(h)$ using orthogonal FDMA/TDMA. System PHY B is a significant extension of PHY A where now each helper implements MU-MIMO to serve its neighboring users and time shares among different possible subsets of served users for zero-forcing beamforming. We describe only PHY B and refer the reader to \cite{bethanabhotla2013joint} for a description of PHY A (which will be used as a term of comparison to show the improvement due to MU-MIMO downlink).

\subsection*{PHY B : MU-MIMO Base Stations} 
In this system, each helper $h$, with a large number of antennas $M$ installed, implements MU-MIMO to serve the users associated to it, i.e., $\Nc(h)$. As a result, helper $h$ can serve simultaneously, in the spatial domain, any subset of size not larger than $\min\{M,|\Nc(h)|\}$ of the clients in $\Nc(h)$. We further assume that each base station performs linear zero-forcing beamforming (LZFBF) to the set of selected users (referred to in the following as ``active users") it serves simultaneously. In addition, each helper $h$ operates in TDMA over all possible active user subsets $\Sm_h \subseteq \Nc(h)$. We use $\Sm_h(\tau)$ to denote the subset that is chosen in OFDM resource block $\tau$, which is different from chunk time $t$ and in fact much smaller than $t$. This assumption is motivated by realistic typical system parameters, where the time and frequency selective wireless channel fading coherence time $\times$ bandwidth product is small with respect to
the number of signal dimensions spanned by the transmission of a video chunk. We let $S_h(\tau)$ to denote the cardinality of  $\Sm_h(\tau)$. Under the assumptions that $M, S_h(\tau) \rightarrow \infty$ with a fixed ratio$\frac{S_h(\tau)}{M} << 1$, random matrix theory results can be invoked to show that for a given choice of subset $\Sm_h(\tau)$ and under the reasonable assumption that the power $P_h$ is equally shared across the user streams in $\Sm_h(\tau)$, the vector of rates achieved by all the users in $\Nc(h)$ is given by
\par \nobreak
{\footnotesize
\begin{align}
 & C_{hu}(\Sm_h(\tau),t) = \notag \\
 &\twopartdef {0} {u \notin \Sm_h(\tau)} {\log\left(1+\frac{g_{hu}(t)(M-S_h(\tau)+1)P_h}{S_h(\tau)\left(1+\sum_{h' \neq h}P_{h'u}g_{h'u}(t)\right)}\right)} {u \in \Sm_h(\tau)}
\label{equal power}
\end{align}
}%
In fact, it is known that the asymptotics kick in very quickly making the rates in~(\ref{equal power}) achievable for practical values of $M$ and $S_h(\tau)$. Notice that the rate expression is independent of the small scale fading coefficients which vary at a faster time scale (i.e. at the scale of OFDM resource blocks). This is because of using a large number of antennas $M$ at the helpers which renders a large $M \times S_h(\tau)$ random channel matrix $\Hm$ of i.i.d complex Gaussian distributed small scale fading coefficients in every OFDM resource block $\tau$. When each helper performs LZFBF in every resource block $\tau$, it turns out that rate expressions involve calculating the reciprocals of the diagonal elements of the inverse Wishart matrix $(\Hm^\mathrm{H}\Hm)^{-1}$.  Due to the large size of the matrix $\Hm$ and under the assumption $\frac{S_h(\tau)}{M} << 1$, random matrix theory results can be invoked to show that the diagonal elements `harden' at a deterministic value. This results in deterministic rate expressions as in (\ref{equal power}) which are independent of $\Hm$ and are just dependent on the large scale path loss coefficients $g_{hu}(t)$ (see \cite{huh2012network} for more details). 
 
Let $\Cm_h(\Sm_h(\tau),t)$ be the $|\Nc(h)|$ dimensional vector whose elements are the rates $C_{hu}(\Sm_h(\tau),t)$ as given in (\ref{equal power}) achieved by every user $ u \in \Nc(h)$ when the helper $h$ beamforms to the active user subset $\Sm_h(\tau)$ in resource block $\tau$. In addition, let $\muv_h(t)$ be the vector obtained by averaging the rate vector $\Cm_h(\Sm_h(\tau),t)$ scheduled by helper $h$ to users in $\Nc(h)$ over all the resource blocks $\tau$ in chunk time $t$. Since we assume that helper $h$ serves its neighboring users $u \in \Nc(h)$ by sharing the resource blocks over all possible active user subsets of $\Nc(h)$, $\muv_h(t)$ is constrained to lie in the LZFBF-achievable region of the underlying MIMO broadcast channel of base station $h$ and users $\Nc(h)$. This yields the transmission rate constraint 
\begin{align}
\muv_h(t) \in \mathrm{coh}\{\Cm_h(\Sm_h,t): \Sm_h \subseteq \Nc(h)\} ~~\forall~~ h \in \Hc
\label{massive-polytope}
\end{align}
where $\mathrm{coh}$ is the short-hand notation for `convex hull'. Note that the above region is a convex polytope in $|\Nc(h)|$ dimensions obtained by taking the convex hull of $2^{|\Nc(h)|}-1$ rate vectors $C_h(\Sm_h,t)$.

In both systems PHY A and PHY B, the slow fading gain $g_{hu}(t)$ models  path loss and shadowing between helper $h$ and user $u$,
and it is assumed to change slowly in time. For a scenario typical of small cell networks, where users are nomadic (e.g., moving at walking speed), 
the slow fading coefficients change on a time-scale of the order of  $10$s (i.e., $\approx 20$ scheduling slots assuming a realistic and common value of $T_{\mathrm{gop}}=0.5$s). This time scale is much slower than
the coherence of the small-scale fading, but is comparable with the duration of the video chunks. Therefore, variations of these coefficients during
a video streaming session (e.g., due to user mobility) are relevant.

We let $\omegav(t)$ denote the network state at time $t$, defined as 
$ \omegav(t) = \left \{ g_{hu}(t), D_{f_u}(\cdot, t), B_{f_u}(\cdot, t) : \forall \; (h,u) \in \Ec \right \}$. 
Let $A_{\omegav(t)}$ be the set of feasible control actions, dependent on the current network state 
$\omegav(t)$, and let $\alpha(t) \in A_{\omegav(t)}$ be a control action, comprising the vector $\muv(t)$ with elements $\mu_{u}(t)$, the quality modes $m_u(t)~\forall~u \in \Uc$ and the vector $\Bm(t)$ with elements $B_{f_u}(m_u(t),t)$. A control policy for the system at hand is a sequence of control actions $\{\alpha(t)\}_{t=0}^{\infty}$ where at each time $t$, $\alpha(t) \in A_{\omegav(t)}$.

\section{Dynamic Streaming Policy Design} \label{sec:numiidstate-dilip}
In the proposed NUM problem, the goal consists of designing a control policy which maximizes a concave utility function of the time averaged {\em video qualities} of the users, subject to keeping the request queues at every user stable. Using the notation $\overline{x} := \lim_{t\rightarrow \infty}\frac{1}{t}\sum_{\tau=0}^{t-1}\E{x(\tau)}$ for the long-term time average expectation of a quantity $x$,  Define $\overline{D}_u:=\lim_{t\rightarrow \infty}\frac{1}{t}\sum_{\tau=0}^{t-1}\E{ D_{f_u}\left(m_u(\tau),\tau\right)}$ and $\overline{Q}_{u} := \lim_{t\rightarrow \infty}\frac{1}{t}\sum_{\tau=0}^{t-1} \E{Q_{u}\left(\tau\right)}$ as the long-term time average of the expected quality level and the expected request queue length respectively at user $u$. Let $\phi_u(\cdot)$ be a concave, continuous, and non-decreasing function defining utility vs. video quality for user $u \in \Uc$. The goal is to solve:
\begin{align}
\textrm{maximize}  & \;\;\; \sum_{u \in \Uc}\phi_u(\overline{D}_u)\label{maxutil}\\
 \textrm{subject to} & \;\;\; \overline{Q}_{u}
< \infty~\forall~ u \in \Uc \label{qstableconst}\\
& \;\;\; \alpha(t) \in A_{\omegav(t)}~\forall~t, \label{feasibleoptions}
\end{align}
where constraint (\ref{qstableconst}) corresponds to the {\it strong stability} condition for all the queues $Q_{u}$ which ensures that all the requested chunks will be eventually delivered. 
Problem (\ref{maxutil}) -- (\ref{feasibleoptions}) can be solved using the stochastic optimization theory of~\cite{neely2010stochastic}.
Since it involves maximizing a {\it function} of time averages, it is convenient to transform it into an equivalent problem that involves 
maximizing a single time average instead of a function of time averages. Then, the {\it drift plus penalty} framework of~\cite{neely2010stochastic} can 
be applied.  This transformation is achieved through the use of auxiliary variables $\gamma_u(t)$ and corresponding virtual queues $\Theta_u(t)$ with buffer evolution: 
\begin{align}
\Theta_u(t+1) = \max{\{\Theta_u(t)+\gamma_u(t)-D_{f_u}(m_u(t),t),0 \}}. \label{virt-update}
\end{align} 
Consider the transformed problem:
\begin{align}
 \textrm{maximize} & \;\;\; \sum_{u \in \Uc}\overline{\phi_u({\gamma}_u)}\label{maxutiltrans}\\
 \textrm{subject to} & \;\;\;  \overline{Q}_{u}
< \infty~\forall~ u \in \Uc \label{qstableconsttrans}\\
& \;\;\; \overline{\gamma}_u \leq \overline{D}_u~\forall~u~\in~\Uc \label{gammaconst}\\
& \;\;\; D_u^{\min} \leq \gamma_u(t) \leq D_u^{\max}~\forall~u~\in~\Uc \label{rectconst}\\
& \;\;\; \alpha(t) \in A_{\omegav(t)}~\forall~t \label{feasibleoptionstrans}
\end{align}

where $D_u^{\max}$ is a uniform upper bound on the maximum quality $D_{f_u}(N_{f_u}, t)$ 
and $D_u^{\min}$ is a lower bound on the minimum quality $D_{f_u}(1,t)$, for all chunk times $t$. 
Notice that constraints~(\ref{gammaconst}) correspond to stability of the virtual queues $\Theta_u$, since $\overline{\gamma}_u$ and $\overline{D}_u$ are the time-averaged arrival rate and the time-averaged service rate for the virtual queue given in (\ref{virt-update}). 

Let ${\bf Q}(t), {\Thetav}(t), \gammav(t)$ and $\Dm(t)$ denote the column vectors with the elements  $Q_{u}(t), \Theta_u(t), \gamma_u(t)$ and $D_{f_u}(m_u(t),t)$  respectively. Let ${\bf G}(t)=\left[ {\bf Q}^\transp(t), \Thetav^\transp(t)\right]^\transp$ be the combined vector of queue backlog vectors
and define the quadratic Lyapunov function $L({\bf G}(t)) := \frac{1}{2} \Gm^\transp (t) \Gm(t)$.  Defining  $\Delta(t):=\E{L({\bf G}(t+1))|{\bf G}(t)}-L({\bf G}(t))$ as the drift at slot $t$, the drift
plus penalty (DPP) policy is designed to solve by observing only the current queue lengths ${\bf Q}(t)$ and the current network state $\omegav(t)$ on each slot $t$ and then choosing $\alpha(t) \in A_{\omegav(t)}$ to minimize a bound on $\Delta(t) -V \sum_{u}\phi_u(\gamma_u(t))$.
Here, $V > 0$ is a control parameter of the policy which affects a utility-backlog tradeoff. 
  It is then easy to show that the resulting policy is given by the minimization, at each chunk time $t$, of the function:
  \par \nobreak
  {\footnotesize
\begin{align}
\underbrace{k{\bf B}^\transp (t) {\bf Q}(t) - \Dm^\transp(t) \Thetav(t)}_{\mbox{\footnotesize congestion control}}&
\;\; - \;\; \underbrace{n{\boldsymbol \mu}^\transp (t){\bf Q}(t)}_{\begin{array}{c} \mbox{\footnotesize transmission scheduling}\end{array}} \notag \\
\;\;& - \;\; 
\underbrace{\left [  V\sum_{u \in \Uc}\phi_u(\gamma_u(t)) -  \gammav^\transp(t)\Thetav(t)  \right ]}_{\mbox{\footnotesize obj. maximization}}
  \label{DDP}
\end{align}
}%
The choice of $m_u(t)~\forall~u \in \Uc$ affects  only the term $k{\bf B}^\transp (t) {\bf Q}(t) -\Dm^\transp(t)\Thetav(t)$, 
the choice of ${\boldsymbol \mu}(t)$ affects only the term $n{\boldsymbol \mu}^\transp (t) {\bf Q}(t)$, 
and the choice of $\gammav(t)$ affects only the term $V\sum_{u \in \Uc}\phi_u(\gamma_u(t)) - \gammav^\transp(t)\Thetav(t)$.
Thus, the overall minimization decomposes into three separate sub-problems. The first two sub-problems have a clear operational meaning, and will be referred to as {\em congestion control} and {\em transmission scheduling}. Congestion control consists of choosing the quality index $m_u(t)$ for the chunk requested at time $t$ and placed into the queue $Q_u$ by every user $u$. Transmission scheduling consists of allocating the channel transmission rates $\mu_{hu}(t)$ for each helper $h$ to its neighboring users $u \in \Nc(h)$. The third sub-problem involves the greedy maximization of each user network utility function with respect to the auxiliary control variables $\gamma_u(t)$. 
\subsection{Congestion Control:Pull Scheme}\label{subsec:no drop-adm control}
The congestion control sub-problem objective function  (see (\ref{DDP})) can be explicitly expressed as
\begin{align*}
\sum_{u \in \Uc} \left \{ k Q_{u}(t)B_{f_u}(m_u(t),t) - \Theta_u(t) D_{f_u}\left(m_u(t),t\right)\right \}.
\end{align*}
The minimization of this quantity decomposes into separate minimizations for each user, namely, 
for each $u \in \Uc$, choose $m_u(t)$ equal to:
\begin{equation} \label{source-coding-rate-decision}
 \arg \min_{m \in \{1, \ldots, N_{f_u}\}}  \left \{ k Q_{u}(t) B_{f_u}(m,t)-\Theta_u(t)D_{f_u}(m,t) \right \}.
\end{equation}
In order to implement this policy, it is sufficient that each user knows the length of its own request queue $Q_u$. Congestion control decisions are decentralized: 
each user $u$ observes the length of its own {\it request queue} $Q_u$  
and requests the $t$-th chunk at a quality level  according to (\ref{source-coding-rate-decision}). 
This policy is reminiscent of the current  adaptive streaming technology for video on demand systems, referred to as DASH (Dynamic Adaptive Streaming over HTTP)~\cite{sanchez2011idash}, 
where the client (user) progressively fetches a video file by downloading successive chunks, 
and makes adaptive decisions on the source encoding quality based on its current knowledge of the congestion of the underlying server-client connection.

\subsection{Transmission Scheduling} \label{subsec:no drop-scheduling}
Transmission scheduling involves maximizing the weighted sum rate 
$\sum_{u \in \Uc} Q_{u}(t) \mu_{u}(t)$ where the weights are the request queue lengths (see (\ref{DDP})).
Under our system assumptions for both PHY A and PHY B, this problem decouples into separate maximizations for each helper. Notice that here, unlike conventional cellular systems, we do not assign a fixed set of users to each helper. In contrast, the helper-user association is dynamic and changes as the users or helpers move around, and results from the transmission scheduling
decision itself. We now describe the algorithm to solve the weighted sum rate maximization at each helper for the system PHY B.
\subsubsection*{Transmission Scheduling for PHY B}
For each $h \in \Hc$, the transmission scheduling problem can be written as the following {\em Linear Program} (LP):
\begin{align}
 \textrm{maximize} & \;\;\; \sum_{u \in \Nc(h)} Q_{u}(t) \mu_{hu}(t)\\
 \textrm{subject to} & \;\;\; \muv_h(t) \in \mathrm{coh}\{\Cm_h(\Sm_h,t): \Sm_h \subseteq \Nc(h)\}.
 \label{coh-const}
\end{align} 
where $\muv_h(t)$ is the $|\Nc(h)|$-dimensional vector of elements of the set $\{\mu_{hu}(t): u \in \Nc(h)\}$. It is well known that the optimal value of an LP is attained at an extreme point/vertex of the polytopal feasible rate region described in (\ref{coh-const}). It is also known that the convex hull of a set of points has a subset of those points as its set of vertices (Prop. 2.2 in \cite{ziegler1995lectures}). Thus, the LP reduces to maximizing the weighted sum rate $\sum_{u \in \Nc(h)} Q_{u}(t) \mu_{hu}(t)$ over the discrete set of points $\{\Cm_h(\Sm_h,t): \Sm_h \subseteq \Nc(h)\}$ of cardinality $2^{|\Nc(h)|}-1$. This means that the subset which maximizes the weighted sum rate is chosen and the same subset is served throughout the entire chunk time slot $t$. This saves a lot of effort in terms of implementation (e.g., overheads in pilot dissemination in a TDD reciprocity-based system) when compared to approaches which switch between different active user subsets every OFDM block $\tau$. 

One can observe from (\ref{equal power}) that when helper $h$ schedules the subset $\Sm_h$ of users for MU-MIMO beamforming, the rate of each user $u \in \Sm_h$ depends only on the cardinality $S_h$ but not on the identity of the members of the subset $\Sm_h$. This implies that for a fixed subset size $S$, the subset of users maximizing the weighted sum rate can be obtained by sorting the users in $\Nc(h)$ according to the weighted rate $Q_u(t)\log\left(1+\frac{g_{hu}(t)(M-S+1)P_h}{S\left(1+\sum_{h' \neq h}P_{h'u}g_{h'u}(t)\right)}\right)$ and choosing greedily the best $S$ users. This `sort+greedy algorithm' is repeated for every subset size $S \in \{1, \ldots, |\Nc(h)|\}$ and the subset which gives the maximum weighted sum rate is chosen. A typical sorting algorithm has complexity $O\left(|\Nc(h)|\log(|\Nc(h)|)\right)$ and since the sorting is repeated for every subset size, our algorithm has complexity $O\left(|\Nc(h)|^2\log(|\Nc(h)|\right)$ which improves upon existing user scheduling algorithms \cite{yoo2006optimality} for the MIMO broadcast channel. Note that the $n\mu_{hu}(t)$ video-encoded bits transmitted by helper $h$ to user $u$ should correspond to the chunks at the head of line of the request queue $Q_u$, assuming the quality chosen by user $u$ in a previous time slot based on the pull scheme~(\ref{source-coding-rate-decision}). Thus, each user $u$ must also broadcast the metadata (chunk number and quality) of the chunks at the head of line along with $Q_u$ to helpers in $\Nc(u)$. In addition, when a user is (possibly) served by multiple helpers in a single slot, the user gets the $n\mu_{u}(t)= \sum_{h \in \Nc(u)}n\mu_{hu}(t)$ video-encoded bits corresponding to the head of line chunks by progressively downloading different PHY layer sub-packets of a chunk (assuming a chunk is segmented into different PHY layer sub-packets for transmission) from different helpers. This would actually require the user to keep track of the PHY layer sub-packets that it has downloaded. However, if intra-session network coding is employed to encode the PHY sub-packets of each chunk, then the user no longer needs to bookkeep at the sub-packet level and just needs to download the required number of linear combinations to decode the chunk. Thus, our approach is ideally suited to an application of distributed storage codes like DRESS~\cite{pawar2011dress} which makes the sub-packet bookkeeping problem much simpler.
 \subsection{Greedy maximization of the network utility function}
Each user $u \in \Uc$ keeps track of $\Theta_u(t)$ and chooses its virtual queue arrival $\gamma_u(t)$ in order to solve:
\par \nobreak
{\footnotesize
\begin{align}
\textrm{maximize}& \;\;\; V\phi_u(\gamma_u(t))-\Theta_u(t)\gamma_u(t) \\
\textrm{subject to }& \;\;\; D_u^{\min} \leq \gamma_u(t) \leq D_u^{\max}.
\end{align}
}%
These decisions push the system to approach the maximum of the network utility function. 
By appropriately choosing the functions $\phi_u$, we can impose some desired notion of fairness.

\section{Algorithm Performance}
\label{nonergodic}
Following in the footsteps of~\cite{bethanabhotla2013joint}, it can be shown that the time average utility achieved by the  policy comes within $O(\frac{1}{V})$ of the utility of a
genie-aided T -slot look ahead policy for any arbitrary sample
path $\omegav(t)$ with a $O(V)$ tradeoff in time averaged backlog. The proof is omitted due to space constraints.

\section{Pre-buffering and re-buffering chunks}  \label{sec:prebuffering} 
The goal here is to determine the delay $T_u$ after which user $u$ should start playback, with respect to the time at which 
the first chunk is requested (beginning of the streaming session).  
We define the size of the playback buffer $\Psi_t$ as the number of playable chunks in the buffer not yet played. Without loss of generality, 
assume again that the streaming session starts at $t = 1$. Then, $\Psi_t$ is recursively given by the updating equation:\footnote{$1\{\Kc\}$ denotes the indicator function
of a condition or event $\Kc$.}
\begin{align}
\Psi_t = \max \left \{ \Psi_{t-1}  -  1\{t > T_u\}, 0 \right \} + |a_t|.
\end{align}
where $|a_t|$ is the number of chunks which are completely downloaded in slot $t$.
Let $A_k$ denote the time slot in which chunk $k$ arrives at the user and let $W_k$ denote the delay with which chunk $k$ is delivered. Note that the longest period during which $\Psi_t$ is not incremented 
is given by the maximum delay to deliver chunks. Thus, each user $u$ needs to adaptively estimate $W_k$ in order to choose $T_u$. In the proposed method,
 at each time $t = 1,2,\ldots$, user $u$ calculates the maximum observed delay $E_t$  in a sliding 
window of size $\Delta$, by letting:
\begin{align}
E_t = \max \{ W_k \; :  ~t-\Delta+1 \leq A_k \leq t \}.
\label{del-window}
\end{align}
Finally, user $u$ starts its playback when $\Psi_t$ crosses the level $\xi E_t$, i.e., $T_u = \min \{ t : ~\Psi_t \geq \xi E_t \}.$
where $\xi$ is a tuning parameter.
If a stall event occurs at time $t$, i.e., $\Psi_t = 0$ for $t > T_u$, the algorithm enters a re-buffering phase in which the same algorithm presented above 
is employed again to determine the new instant $t + T_u + 1$ at which playback is restarted. 

\section{Numerical Experiment} \label{sec:simul-dilip}
\begin{figure}[t]
\subfloat[]{
\includegraphics[width = 40 mm, height = 30mm]{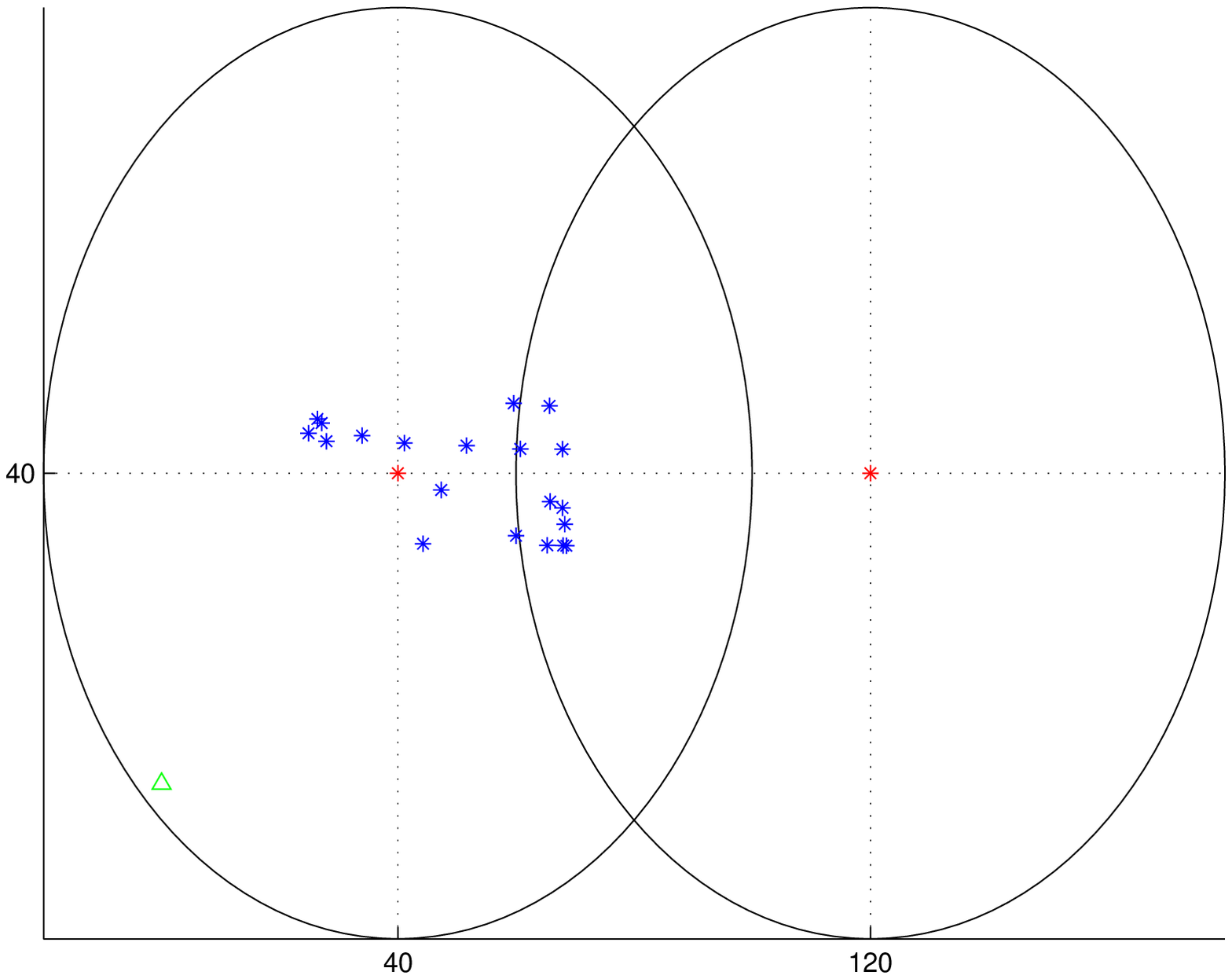}
\label{topology}
}
\subfloat[CDF of avg. video quality]{
\includegraphics[width = 40 mm, height = 30mm]{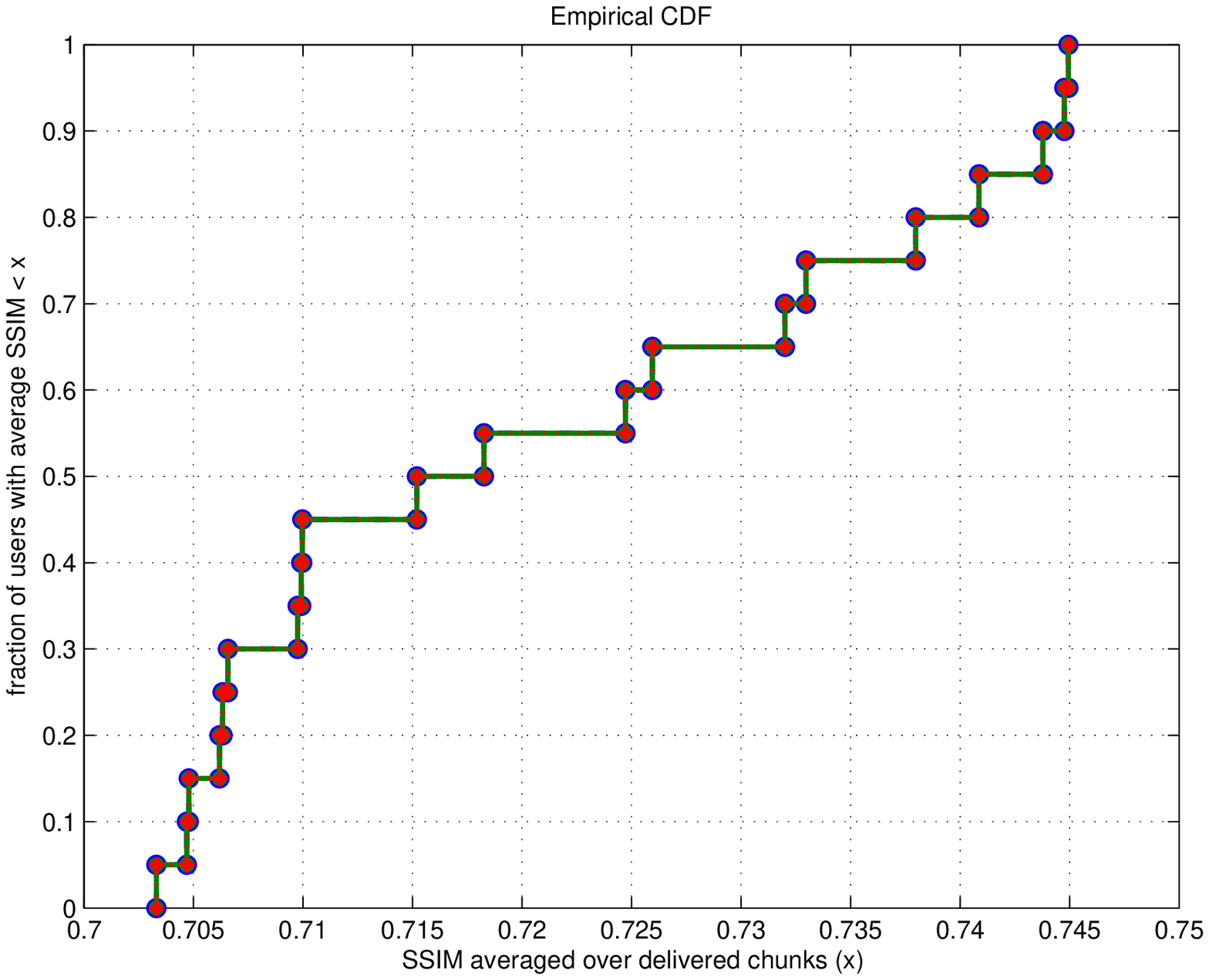}
\label{ssim-cdf-mimo}
}\\
\subfloat[CDF of pre buffering time]{
\includegraphics[width =  40 mm, height = 30mm]{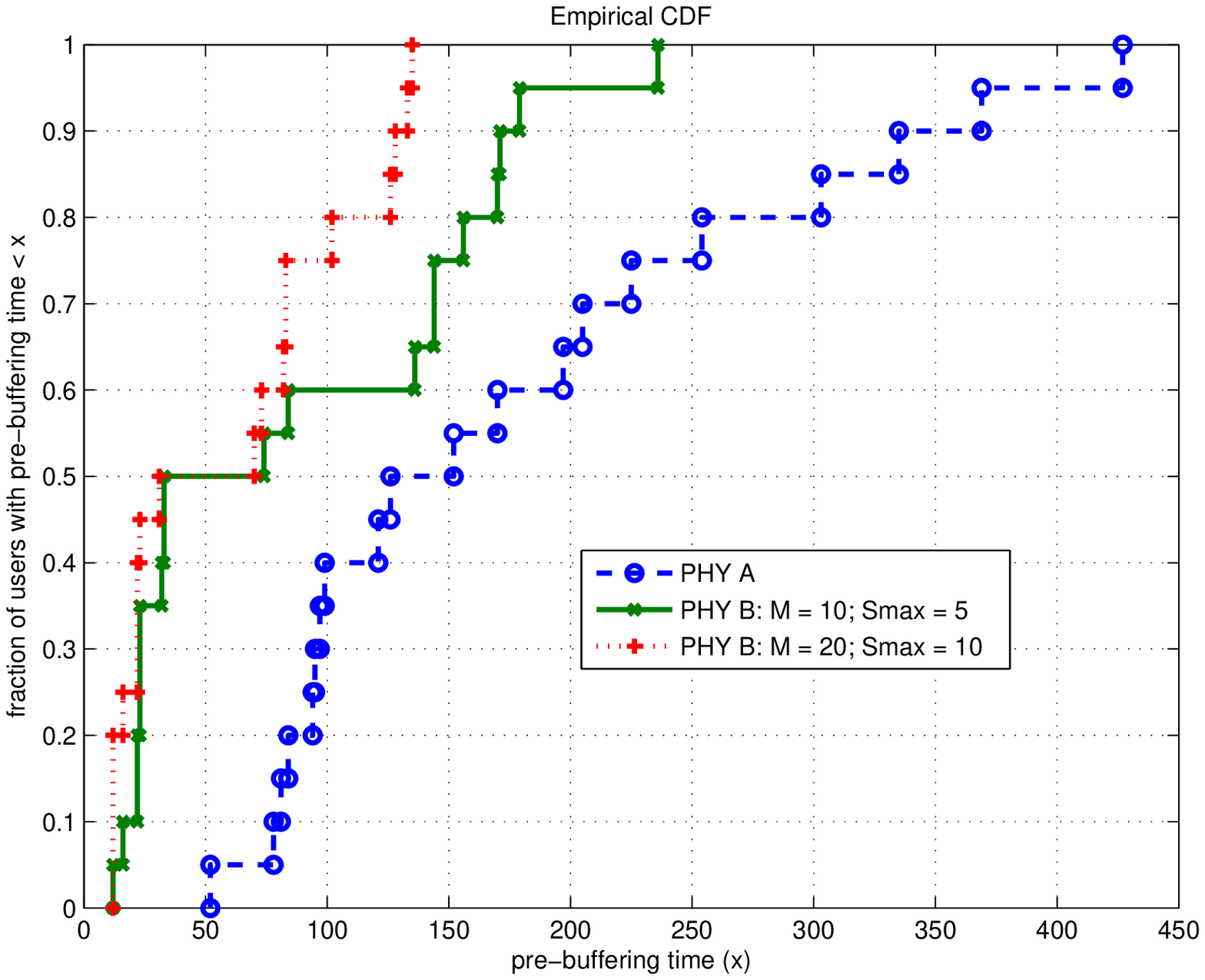}
\label{prerebuff-cdf-mimo}
}
\subfloat[CDF of rebuffing percentage]{
\includegraphics[width = 40 mm, height = 30mm]{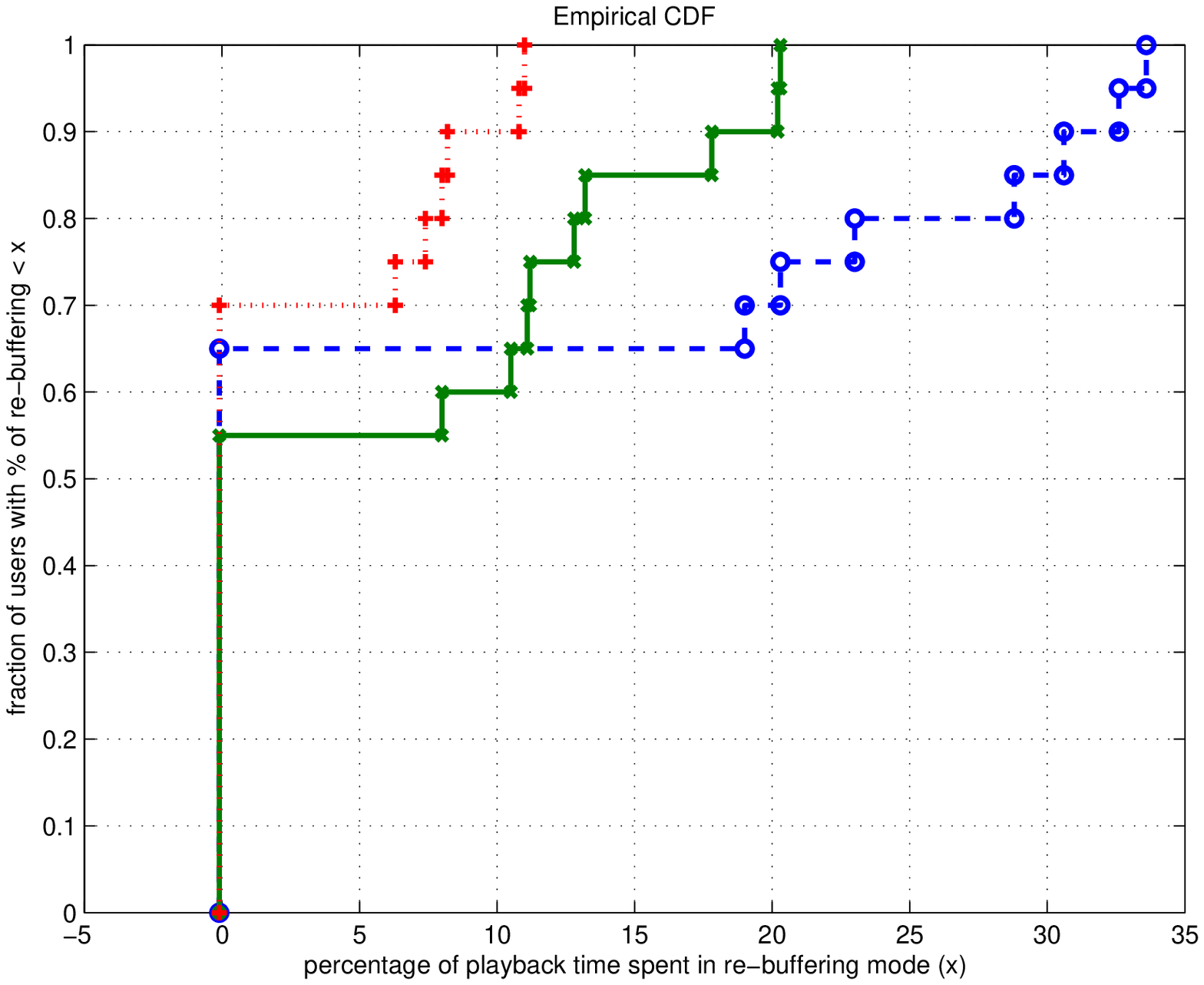}
\label{rebuff-cdf-mimo}
}\\

\end{figure}

We consider a simple topology with $2$ helpers and $20$ users as shown in Figure (\ref{topology}). The parameters of the simulation are chosen to be identical to those chosen in the experiments in~\cite{bethanabhotla2013joint}. We choose the utility function $\Phi_u(\cdot) = \log(\cdot)~\forall~u \in \Uc$ to impose proportional fairness. We use the same video file in~\cite{bethanabhotla2013joint} with VBR encoded chunks and with video quality measured using the SSIM index~\cite{wang2004image}. We simulate the algorithms designed above and plot the CDF over the user population of video streaming QoE metrics. We observe that the average video quality performance is the same in all systems but there is a significant improvement of MU-MIMO PHY B over PHY A in terms of pre-buffering and re-buffering times. The blue curves correspond to system PHY A. The green curves correspond to PHY B with number of antennas $M =10$ and maximum active user subset size $S_{\max} = 5$. The red curves correspond to PHY B with $M=20$ and $S_{\max}=10$. 

\bibliographystyle{IEEEtran}
\bibliography{dilip-ref}

\begin{thebibliography}{10}
\providecommand{\url}[1]{#1}
\csname url@samestyle\endcsname
\providecommand{\newblock}{\relax}
\providecommand{\bibinfo}[2]{#2}
\providecommand{\BIBentrySTDinterwordspacing}{\spaceskip=0pt\relax}
\providecommand{\BIBentryALTinterwordstretchfactor}{4}
\providecommand{\BIBentryALTinterwordspacing}{\spaceskip=\fontdimen2\font plus
\BIBentryALTinterwordstretchfactor\fontdimen3\font minus
  \fontdimen4\font\relax}
\providecommand{\BIBforeignlanguage}[2]{{%
\expandafter\ifx\csname l@#1\endcsname\relax
\typeout{** WARNING: IEEEtran.bst: No hyphenation pattern has been}%
\typeout{** loaded for the language `#1'. Using the pattern for}%
\typeout{** the default language instead.}%
\else
\language=\csname l@#1\endcsname
\fi
#2}}
\providecommand{\BIBdecl}{\relax}
\BIBdecl

\bibitem{bethanabhotla2013joint}
D.~Bethanabhotla, G.~Caire, and M.~J. Neely, ``Joint transmission scheduling
  and congestion control for adaptive video streaming in small-cell networks,''
  \emph{arXiv preprint arXiv:1304.8083}, 2013.

\bibitem{cisco66}
Cisco, ``The zettabyte era-trends and analysis,'' 2013.

\bibitem{sesia-LTE}
S.~Sesia, I.~Toufik, and M.~Baker, \emph{{LTE}: the Long Term Evolution-From
  theory to practice}.\hskip 1em plus 0.5em minus 0.4em\relax Wiley, 2009.

\bibitem{hoydis2011massive}
J.~Hoydis, S.~Ten~Brink, and M.~Debbah, ``Massive {MIMO}: How many antennas do
  we need?'' in \emph{{2011 49th Annual Allerton Conference on Communication,
  Control, and Computing (Allerton)}}.\hskip 1em plus 0.5em minus 0.4em\relax
  IEEE, 2011, pp. 545--550.

\bibitem{hoydis2011green}
J.~Hoydis, M.~Kobayashi, and M.~Debbah, ``Green small-cell networks,''
  \emph{Vehicular Technology Magazine, IEEE}, vol.~6, no.~1, pp. 37--43, 2011.

\bibitem{802.11ac}
E.~H. Ong, J.~Kneckt, O.~Alanen, Z.~Chang, T.~Huovinen, and T.~Nihtila, ``{IEEE
  802.11 ac}: Enhancements for very high throughput {WLAN}s,'' in \emph{{2011
  IEEE 22nd International Symposium on Personal Indoor and Mobile Radio
  Communications (PIMRC)}}.\hskip 1em plus 0.5em minus 0.4em\relax IEEE, 2011,
  pp. 849--853.

\bibitem{yi2008stochastic}
Y.~Yi and M.~Chiang, ``Stochastic network utility maximisation-a tribute to
  {K}elly's paper published in this journal a decade ago,'' \emph{European
  Transactions on Telecommunications}, vol.~19, no.~4, pp. 421--442, 2008.

\bibitem{neely2010stochastic}
M.~Neely, ``Stochastic network optimization with application to communication
  and queueing systems,'' \emph{Synthesis Lectures on Communication Networks},
  vol.~3, no.~1, pp. 1--211, 2010.

\bibitem{sanchez2011idash}
Y.~S{\'a}nchez, T.~Schierl, C.~Hellge, T.~Wiegand, D.~Hong, D.~De~Vleeschauwer,
  W.~Van~Leekwijck, and Y.~Lelouedec, ``i{D}{A}{S}{H}: improved dynamic
  adaptive streaming over http using scalable video coding,'' in \emph{ACM
  Multimedia Systems Conference (MMSys)}, 2011, pp. 23--25.

\bibitem{begen2011watching}
A.~Begen, T.~Akgul, and M.~Baugher, ``Watching video over the web: Part 1:
  Streaming protocols,'' \emph{Internet Computing, IEEE}, vol.~15, no.~2, pp.
  54--63, 2011.

\bibitem{ortega2000variable}
A.~Ortega, ``Variable bit-rate video coding,'' \emph{Compressed Video over
  Networks}, pp. 343--382, 2000.

\bibitem{wang2004image}
Z.~Wang, A.~C. Bovik, H.~R. Sheikh, and E.~P. Simoncelli, ``Image quality
  assessment: From error visibility to structural similarity,'' \emph{Image
  Processing, IEEE Transactions on}, vol.~13, no.~4, pp. 600--612, 2004.

\bibitem{huh2012network}
H.~Huh, A.~M. Tulino, and G.~Caire, ``Network mimo with linear zero-forcing
  beamforming: Large system analysis, impact of channel estimation, and
  reduced-complexity scheduling,'' \emph{Information Theory, IEEE Transactions
  on}, vol.~58, no.~5, pp. 2911--2934, 2012.

\bibitem{ziegler1995lectures}
G.~M. Ziegler, \emph{Lectures on polytopes}.\hskip 1em plus 0.5em minus
  0.4em\relax Springer, 1995, vol. 152.

\bibitem{yoo2006optimality}
T.~Yoo and A.~Goldsmith, ``On the optimality of multiantenna broadcast
  scheduling using zero-forcing beamforming,'' \emph{Selected Areas in
  Communications, IEEE Journal on}, vol.~24, no.~3, pp. 528--541, 2006.

\bibitem{pawar2011dress}
S.~Pawar, N.~Noorshams, S.~El~Rouayheb, and K.~Ramchandran, ``Dress codes for
  the storage cloud: Simple randomized constructions,'' in \emph{Information
  Theory Proceedings (ISIT), 2011 IEEE International Symposium on}.\hskip 1em
  plus 0.5em minus 0.4em\relax IEEE, 2011, pp. 2338--2342.

\end{thebibliography}

\end{document}